\documentclass[aps,prl,reprint,groupedaddress,showkeys]{revtex4-2}
\usepackage{graphicx}
\usepackage{amsmath}

\begin{document}

%\preprint{}

\title{Broadband and intense sound transmission loss by a coupled-resonance acoustic metamaterial}


\author{David Roca}
\email[]{droca@cimne.upc.edu}
%\thanks{}
\affiliation{
	Centre Internacional de M\`{e}todes Num\`{e}rics en Enginyeria (CIMNE), Universitat Polit\`{e}cnica de Catalunya, Barcelona 08034, Spain
}

\author{Mahmoud I. Hussein}
\email[]{mih@colorado.edu}
%\thanks{}
\affiliation{
	Ann and HJ Smead Department of Aerospace Engineering Sciences, University of Colorado Boulder, Boulder, Colorado 80303, USA
}
\affiliation{
	Department of Physics, University of Colorado Boulder, Boulder, Colorado 80302, USA
}


\date{\today}

\begin{abstract}
The advent of acoustic metamaterials opened up a new frontier in the control of sound transmission. A key limitation, however, is that an acoustic metamaterial based on a single local resonator in the unit cell produces a restricted narrow-band attenuation peak; and when multiple local resonators are used the emerging attenuation peaks$–$while numerous$–$are each still narrow and separated by pass bands. Here, we present a new acoustic metamaterial concept that yields a sound transmission loss through two antiresonances$-$in a single band gap$-$that are fully coupled and hence provide a broadband attenuation range; this is in addition to delivering an isolation intensity that exceeds 90 decibels for both peaks. The underlying coupled resonance mechanism is realized in the form of a single-panel, single-material pillared plate structure with internal contiguous holes$-$a practical configuration that lends itself to design adjustments and optimization for a frequency range of interest, down to sub-kilohertz, and to mass fabrication.        

%Sound transmission loss (STL) peaks can be achieved with acoustic metamaterials by matching the desired frequency range of attenuation with the corresponding bandgap regions in their associated unit cell's dispersion characteristics. Since these bandgaps are typically produced by local resonance effects, they tend to be narrowband, limiting the STL performance of the acoustic metamaterial for practical applications. Here, we show a mechanism to broaden the range of effective STL attenuation using coupled internal resonators in a doubly periodic acoustic metamaterial supercell. The improved STL properties obtained are demonstrated in a practical realization consisting on a pillared plate with holes, exhibiting the coupling mechanism. A parametric study shows that there is room for optimizing the proposed design performance to obtain broadband STL response at lower frequency ranges.
\end{abstract}


\maketitle

\begin{figure}
	\includegraphics{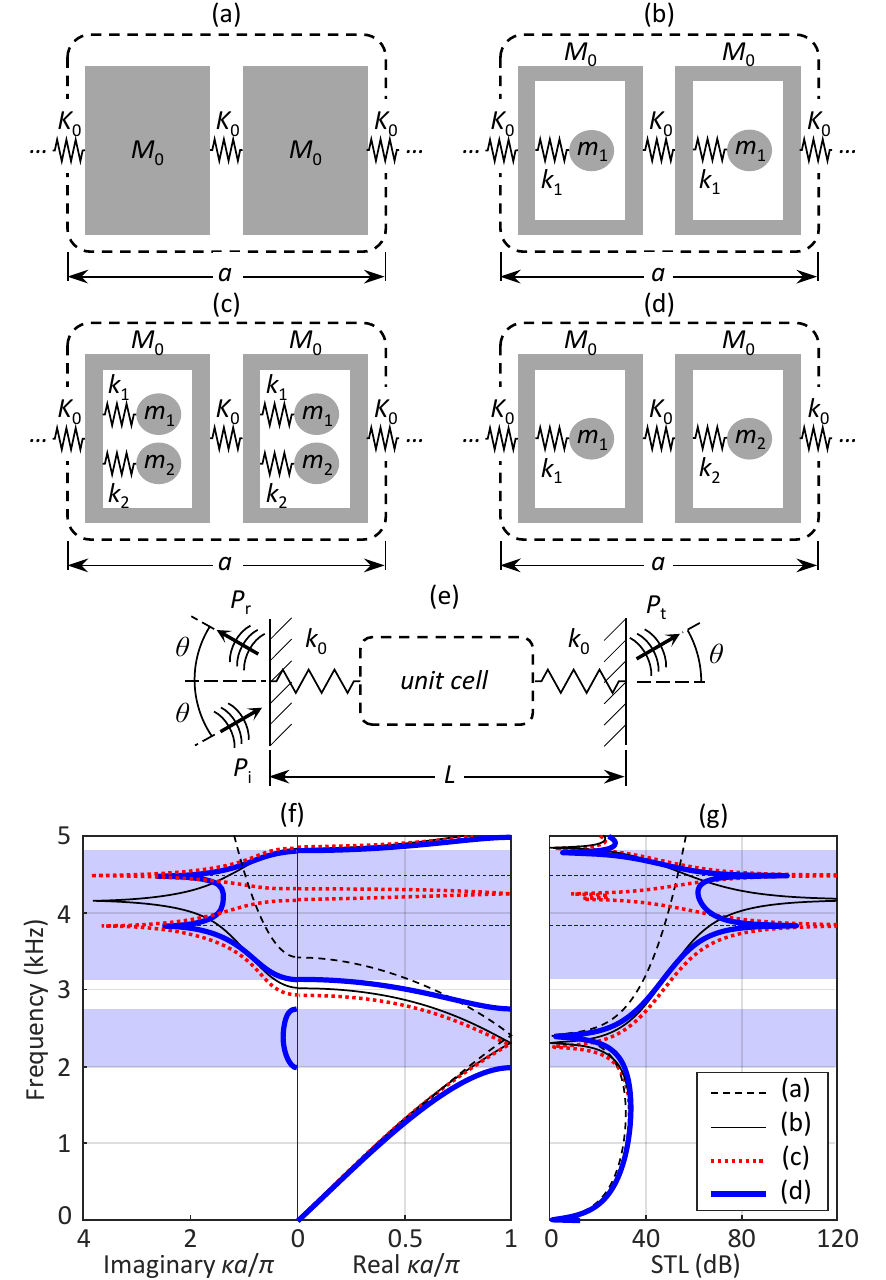}%
	\caption{Supercell configurations for four different spring-mass models: (a) Identical mass units (phononic crystal), (b) identical mass-in-mass units with a single resonator (classical acoustic metamaterial), (c) identical mass-in-mass units with two different resonators, and (d)  different single-resonator mass-in-mass units in a dual periodicity configuration.~(e) Panel model considered for calculating the STL for acoustic plane waves at an angle of incidence $\theta$.~The panel is surrounded by air at both sides.~A pressure wave with amplitude $P_{\text{i}}$ is incident$-$a portion is transmitted, $P_{\text{t}}$, while the remaining portion is reflected back, $P_{\text{r}}$.~This yields $\text{STL} = 10\log_{10} |P_{\text{i}}/P_{\text{t}}|^2$ (in decibels).~(f) Dispersion curves for each supercell model (shaded regions mark frequency band gaps).~(g) STL curves for each supercell model for $\theta=45$\textdegree.~The parameters have been chosen to provide an STL spectrum similar to that of the HP${}^2$ configuration in Fig.~\ref{Fig3}.~Since only one unit cell is considered in the panel model, $L=a$ for all four spring-mass models. \label{Fig1}}
\end{figure}

A long-standing problem in acoustics is the attenuation of sound in the low-mid range of the frequency spectrum (below 5~kHz).~Use of conventional materials for this purpose is restricted by the classical mass law which states that the level of sound transmission through a solid panel in air is proportional to the inverse of the product $\omega \rho t$, where $\omega$ is the frequency and $\rho$ and $t$ are the density and thickness of the solid panel, respectively \cite{Berger1911, Heckl1981}.~Thus, an insulation panel in a building or a vehicle has to be both thick$-$often beyond a practical range$-$and composed of a dense material to provide an intense level of sound transmission loss (STL) \cite{Osipov1997}.~Modest improvements in STL performance have been realized by incremental techniques such as considering double or multi-layered panels separated by air to benefit from modal interactions \cite{Beranek1949,Mulholland1967,Legault2010} and sandwich panels with a viscoelastic material in between to generate increased dissipation \cite{Dym1974}.~Perforated plates \cite{Uris2001} or other forms of panel-shaped sonic crystals \cite{Garcia-Chocano2012, Morandi2016} have also been investigated utilising in-plane Bragg scattering effects, but these require the unit cell planar size to be on the order of the wavelength of the incident sound waves in air, which are impractically large$-$on the order of decimeters.  \\ 
\indent A turning point in the field emerged when the fundamental limitations of the classical mass law were overcome with the introduction of \it acoustic metamaterials \rm by Sheng and co-workers in 2000~\cite{Liu2000}.~An acoustic metamaterial comprises local resonators that are intrinsically either embedded in a three-dimensional (3D) elastic medium$-$as demonstrated in Ref.~\cite{Liu2000}$-$or attached to a 2D elastic medium such as a taut membrane \cite{Yang2008, Yang2010} or a plate~\cite{Pennec2008,Wu2008,Assouar2016}.~In an acoustic/elastic metamterial, local resonances hybridize with dispersion curves of the underlying medium causing the opening of band gaps that may be tuned to appear deep into the subwavelength regime.~The study of acoustic, and elastic, metamaterials has witnessed expansive growth, touching on a wide range of applications, over the past two decades~\cite{Hussein2014, Cummer2016, Assouar2018}. \\
\indent Acoustic metamaterials in the form of stretched membranes with attached disc-shaped resonators have been employed for sound absorption \cite{Yang2008, Yang2010}. These materials, while effective in the sub-kilohertz frequency range$-$which is the target range of the human auditory perception \cite{Rossing2007}$-$require tension and are thus non-load bearing without a supporting frame surrounding each unit cell.~In contrast, pillared plates are load bearing with some designs have been shown to exhibit STL levels exceeding 60 decibels$-$for an aluminum plate of thickness 0.5 mm and rubber pillars with a loss factor of 0.01~\cite{Assouar2016}.~However, being a conventional acoustic metamaterial, the STL peaks of a pillared plate are isolated and narrow in their bandwidth$-$which limits its effectiveness for practical application. Introducing pillars with different natural frequencies in the unit cell opens up multiple band gaps, which enables multiple, closely packed peaks in the STL spectrum \cite{Xiao2012}; however, these peaks are each still narrow and separated by transmission bands.~Several proposals have been presented to widen the band gaps of locally resonant acoustic/elastic metamaterials (see, e.g., Refs. \cite{Bilal2013,Matsuki2014,Krushynska2014,Roca2019}); however, all these strategies are constrained by the fundamental limitation of a single, narrow attenuation peak in the band gap as directly observed in the imaginary wave vector portion of the dispersion band structure \cite{Frazier2016}.  \\
\indent In this letter, we present a new acoustic metamaterial concept$-$and associated  geometrical architecture$-$for a loading-bearing single-panel, single-material wall that yields a unique physical response in the STL spectrum with characteristics only qualitatively attainable by membrane-based double panels comprising multiple materials.~Our proposed configuration, termed \it hollow-plate/hollow-pillar \rm (HP$^2$), generates two coupled STL peaks, at significantly elevated decibel levels, enabling a broader bandwidth coverage compared to a corresponding conventional pillared plate with its characteristic isolated single peaks in the STL spectrum.~The realization of coupled double peaks within a single band gap in a locally resonant elastic metamaterial has recently been demonstrated using the notion of \it dual periodicity \rm by~\citet{Gao2020} and Li and Wang \cite{Li2020}.~In Fig.~\ref{Fig1}, we demonstrate the effects of dual periodicity in a simple lumped-parameter spring-mass STL model where two distinct mass-in-mass \cite{Huang2009} units are connected (see Fig.~\ref{Fig1}d) and repeated periodically as a larger unit (supercell).~The dispersion curves for model (d), shown in Fig.~\ref{Fig1}f, exhibit two band gaps$-$the first (between 2000 and 2750 Hz) is attributed to Bragg scattering effects while the second is produced by a coupling of the two different local resonances of each mass-in-mass in the supercell (occurring at 3840 and 4490, respectively).~This coupling, attributed to the dual periodicity, has an effect on the corresponding STL curve in Fig.~\ref{Fig1}g where the two peaks are joined in a continuous attenuation band with a broad bandwidth.~For comparison, the statically equivalent arrangement of springs and masses in model (c) also produces two STL peaks at the same frequencies, but separated by an STL antipeak (due to the presence of a transmission band in between, as shown in Fig.~\ref{Fig1}f).~Results for a statically equivalent phononic crystal and acoustic metamaterial with single periodicity and only one resonator in the unit cell (models (a) and (d), respectively) are also shown for comparison. \\
\begin{figure}
\includegraphics{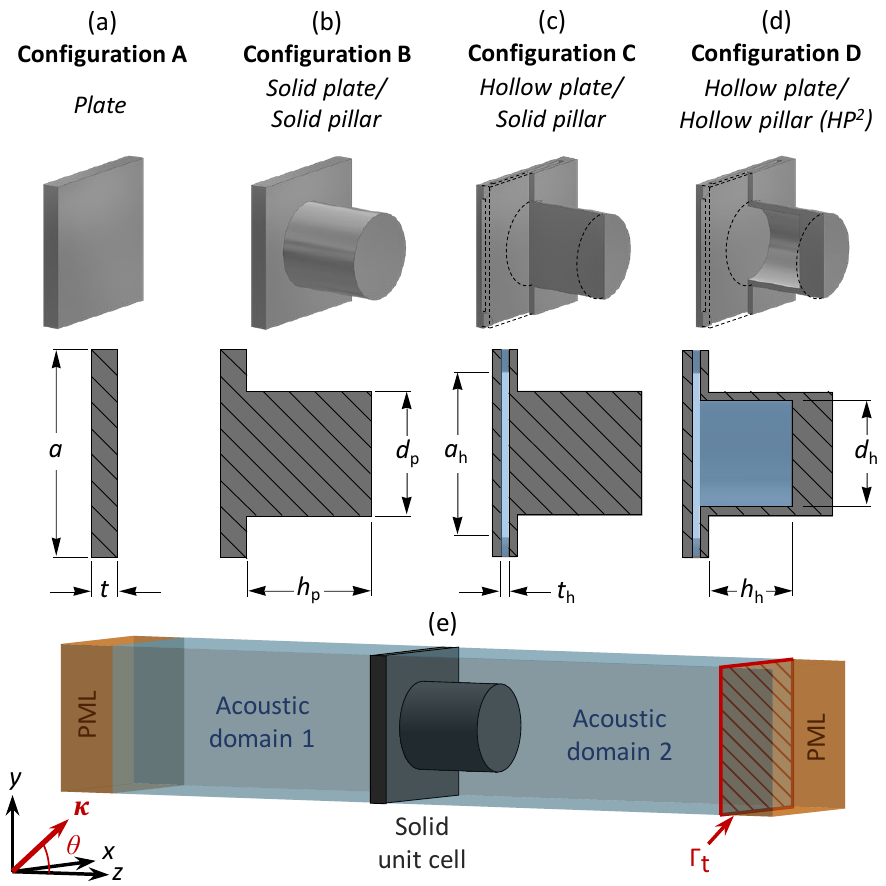}%
\caption{3D unit-cell models for a selection of panel configurations: (a) Constant-thickness plate, (b) standard all-solid pillared plate, (c) hollow-plate/solid-pillar and (d) hollow-plate/hollow-pillar (HP${}^2$). Side-cut section views showing the relevant geometric parameters.~(e) System model for computing STL for each panel configuration.~Two perfectly matched layers (PMLs) are prescribed on each end of the acoustic (air) domains to simulate an infinite medium in the $z$-direction.~Bloch boundary conditions are imposed on both the solid and acoustic domains' boundaries, allowing restriction of the simulations to a single unit cell representing an infinite periodic repetition of the model in both $x$ and $y$-directions.~An acoustic plane wave with an amplitude $P_{\text{i}}=0.02$ Pa (equivalent to 60 dB of sound pressure level) and an incidence angle $\theta$ is imposed onto acoustic domain 1 as a background pressure field.~The scattered pressure field is solved in both acoustic domains along with the displacement field in the solid domain.~The average pressure field amplitude over the transmission surface $\Gamma_{\text{t}}$ (i.e., $P_{\text{t}} = \int_{\Gamma_{\text{t}}}|p_2|d\Gamma$) is then computed and fed into the STL formula provided in the caption of Fig.~\ref{Fig1}. \label{Fig2}}
\end{figure}
\begin{figure}
\includegraphics{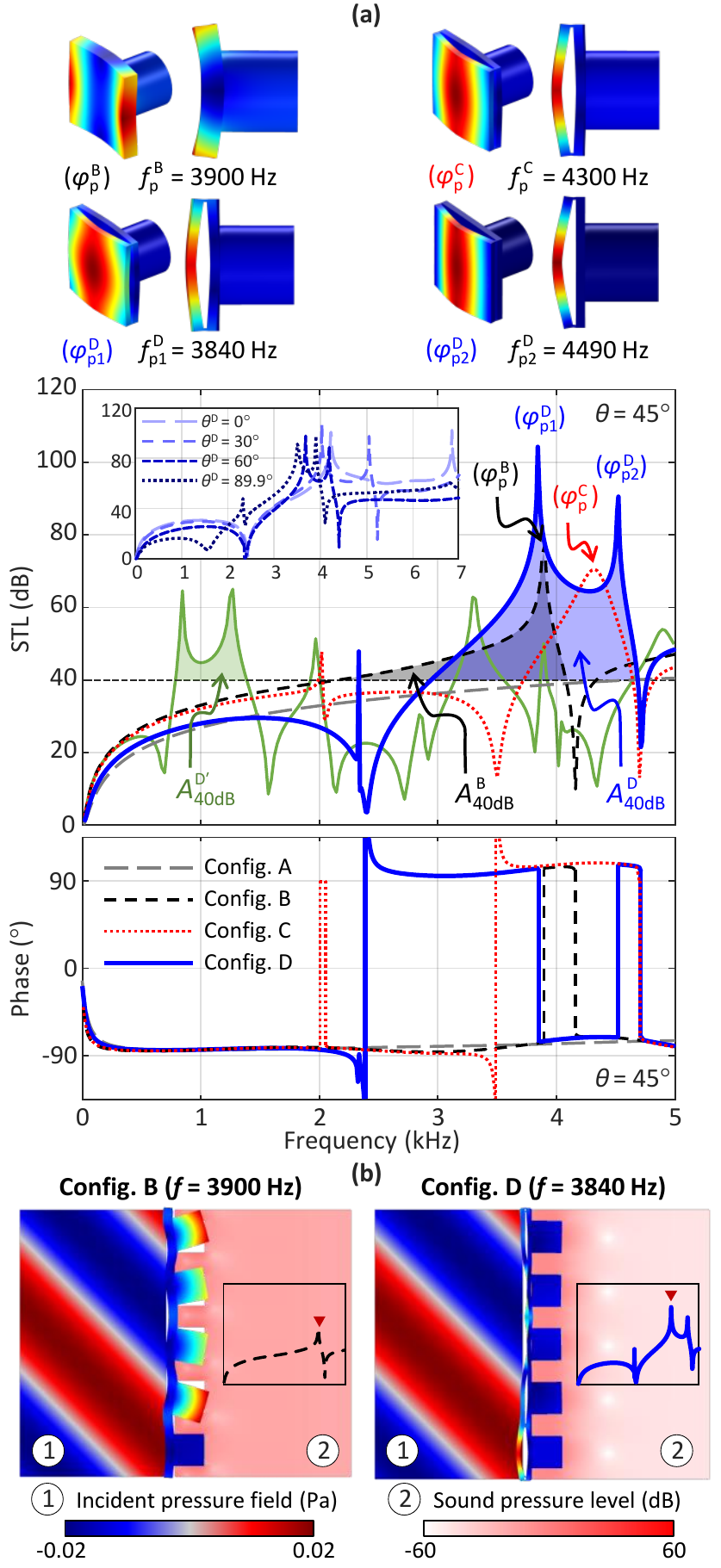}%
\caption{(a) STL curves and corresponding phase shift curves.~The main plots show results for an angle of incidence of $\theta=45\text{\textdegree}$ for all panel configurations, while the inset details the STL response of the HP${}^2$ case (Config.~D) at different angles of incidence.~The mode shapes corresponding to several peaks in the STL curves are shown at the top marked by their frequencies.~The green curve in the background of the main STL plot shows the response of a modified HP${}^2$ panel configuration with added thin metal discs that also features double peaks but centered around 1~kHz (See SI section).~The shaded regions in gray, blue, and green, labelled $A_{40\text{dB}}^{\text{B}}$,  $A_{40\text{dB}}^{\text{D}}$, and $A_{40\text{dB}}^{\text{D$^\prime$}}$, respectively, define the area (in dB$\cdot$kHz) below the corresponding curve and above a 40 dB baseline. (b) Pressure fields for Configs. B (left) and D (right).\label{Fig3}}
\end{figure}
\begin{figure*}
\includegraphics{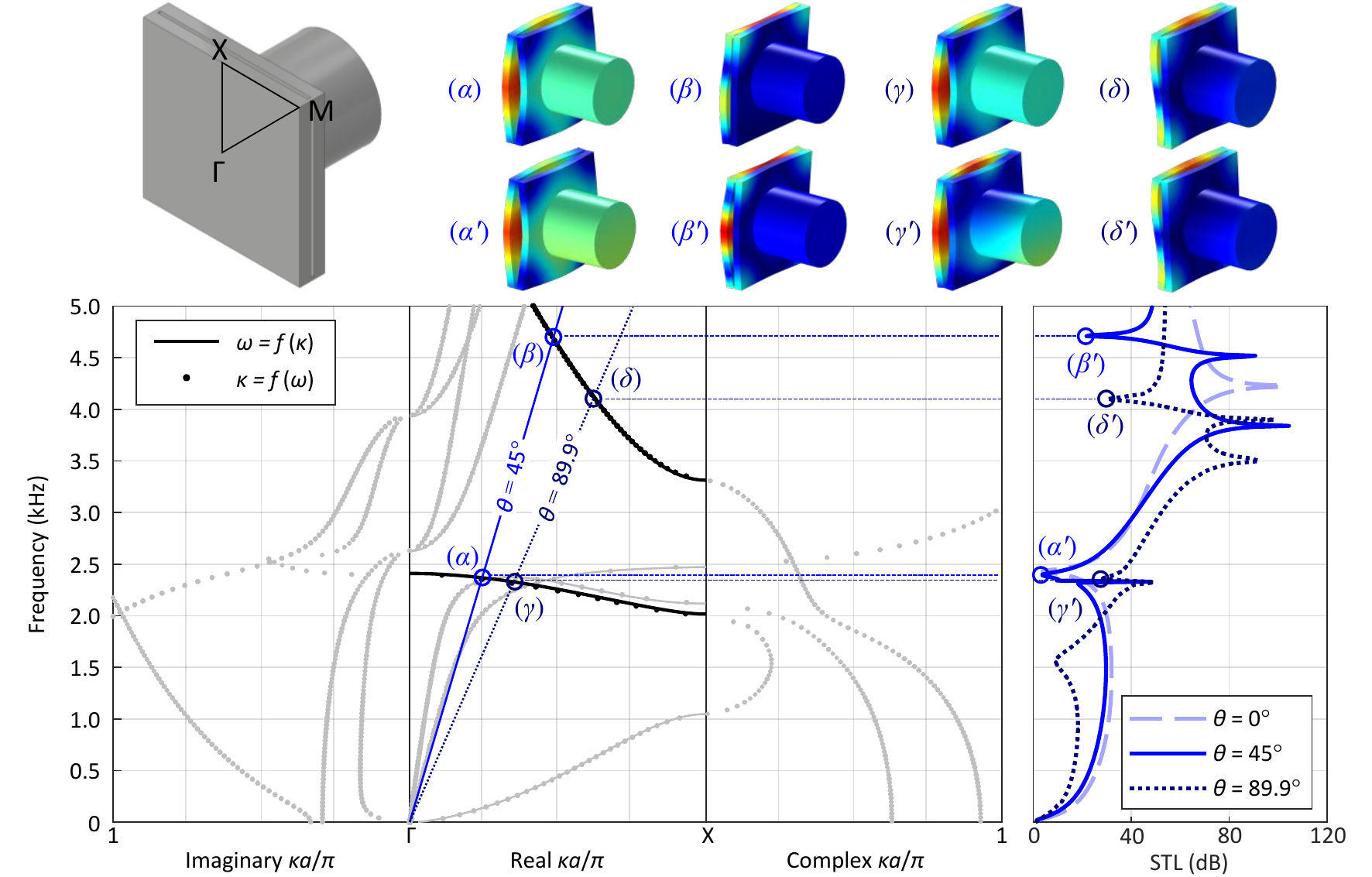}%
\caption{Complex band structure of the HP${}^2$ panel configuration in the $\Gamma \text{X}$ direction (left) and corresponding STL curves for different angles of incidence (right).~The dispersion results have been obtained by FEM analysis following a frequency as a $\kappa(\omega)$ formulation (dots) to access the complex wave-vector space ($\kappa$: wavenumber, $\omega$: frequency).~For the real part, results using a standard $\omega(\kappa)$ formulation (solid lines) are also given for reference.~The blue and red oblique dashed lines in the real part of the dispersion plot correspond to the sound lines for air at the indicated angles of incidence.~The mode shapes corresponding to their intersections with the highlighted branches in the dispersion plot are shown at the top, alongside the mode shapes associated with the corresponding STL curves' antipeaks. \label{Fig4}}
\end{figure*}
\begin{figure*}
	\includegraphics{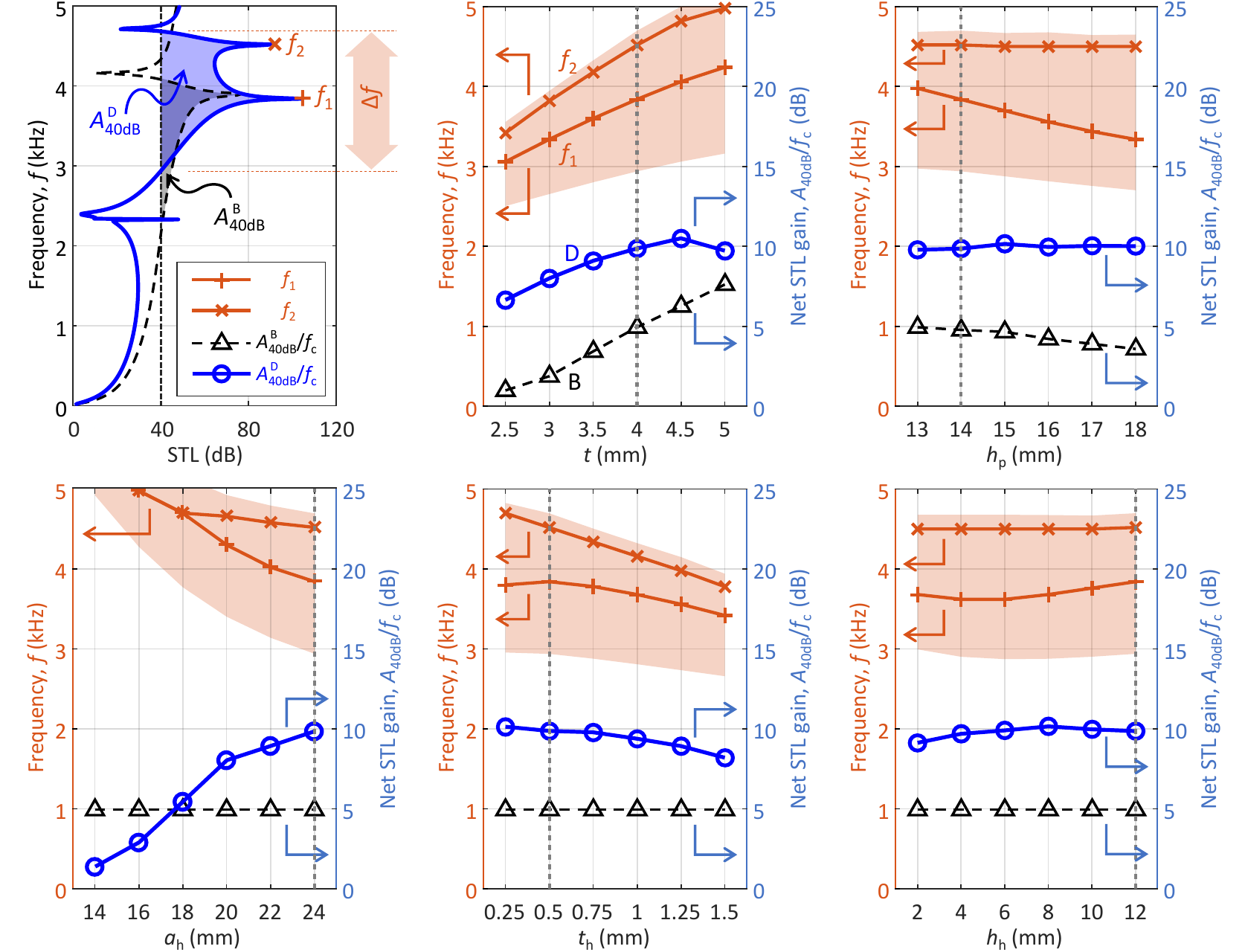}%
	\caption{Plot on the top left corner shows the STL curves for Config.~B (black dashed line) and Config.~D (blue solid line).~For the HP${}^2$ configuration (Config.~D), the coupled peaks frequencies are defined by $f_1$ and $f_2$, and the corresponding attenuation bandwidth above 40 dB is denoted by $\Delta f$.~Remaining plots show the influence of different geometric parameters (as defined in Fig.~\ref{Fig2}) on the net STL gain for both Config.~B and Config.~D (defined as their corresponding $A_{40\text{dB}}$ divided by the central frequency $f_{\text{c}} = [f_1+f_2]/2$).~These plots also show for Config.~D the corresponding attenuation band above 40 dB (depicted by the shaded areas) as well as the peak frequencies in terms of the geometric parameters.~The black dashed vertical lines indicate the results for the nominal HP$^2$ configuration examined in top-left plot and in Figs.~\ref{Fig3} and~\ref{Fig4}.	\label{Fig5}}
\end{figure*}
\indent In Figure~\ref{Fig2}, we present an example configuration of our proposed HP$^2$ panel. Unlike the system displayed in Fig.~\ref{Fig1}, here we only have a single periodicity, i.e., no distinct mass-in-mass units are needed.~Instead, we realize the coupled double STL peak response by introducing two contiguous holes, one within the body of a plate portion and the other within the body of an attached pillar$-$hence the HP$^2$ terminology.~The holes are filled with air, however alternatively they could be in vacuum as demonstrated in the Supplementary Material (SM) section.~The connection between the holes enable an elastic modal interaction that yields two coupled peaks in the STL spectrum as shown in Fig.~\ref{Fig3}. The panels are made out of the polymeric material ABS (amenable to 3D-printing) for which the density is 1050 kg/m${}^3$, the Young's modulus is 1627 MPa, and the Poisson's ratio is 0.35.\footnote{These properties are similar to those of plaster, a practical material used for walls in buildings.}~We assume no damping, thus lower STL values are expected in practice~\cite{VanBelle2019}. However, as shown the SM section, results for experimental values of loss factors for ABS indicate that the differences are very small, with peak levels still being above 90 dB.~For the geometric dimensions, the following set of practically suitable parameters (defined in Fig.~\ref{Fig2}) are chosen for a nominal configuration: $a=25$~mm, $t=4$~mm, $h_{\text{p}}=14$~mm, $d_{\text{p}}=15$~mm, $a_{\text{h}}=24$~mm, $t_{\text{h}}=0.5$~mm, $h_{\text{h}}=12$~mm and $d_{\text{h}}=11.5$~mm. \\
\indent The STL spectrum for the HP${}^2$ panel (Config.~D) is given in Fig.~\ref{Fig3} showing an elevated coupled double peak.~This STL profile is far superior, in both amplitude and bandwidth, than that of the same configuration without the internal holes (Config.~B); the latter exhibits a conventional single resonance peak.~Results for a corresponding single plate without pillars or holes (Config.~A) allows us to establish a baseline value of 40 dB that we use to define the \textit{net STL gain}, a performance metric that quantifies the area below the STL curve and above this 40 dB baseline (see Fig.~\ref{Fig3}) divided by the central frequency (average frequency between the two peaks).~The net STL gain for the HP${}^2$ configuration is shown to be roughly double that of Config.~B.~Moreover, as shown in the inset, the coupled-resonance phenomenon exists for the full range of angles of incidence between 0\textdegree and 90\textdegree.~The distance between the peaks (and hence the net attenuation bandwidth) increases upon approaching normal incidence ($\theta=0\text{\textdegree})$ and the frequency range decreases for angles closer to 90\textdegree.\\
\indent The nature of the double-peak formation mechanism is elucidated by correlating the STL response with the elastic band structure for in-plane wave propagation in the HP${}^2$ panel.~Figure~\ref{Fig4} demonstrates this correlation along the $\Gamma\text{X}$ wave vector path in the Brillouin zone.~It is seen that the coincidence frequencies corresponding to the STL antipeaks surrounding the double peaks match with the intersections of the associated sound lines with two branches of the dispersion curves; this is consistent with theory as demonstrated for conventional acoustic panels~\cite{Wang2005}.~It is observed that the modes corresponding to these points also match with the associated modes of the STL antipeaks. \\
\indent In Figure~\ref{Fig5}, we examine the influence on the net STL gain of varying certain HP${}^2$ geometric parameters with respect to the nominal case considered above.~Results for the HP${}^2$ configuration show a sizable improvement over the corresponding configuration without holes, with practically a doubling of the net STL gain in most cases.~Increasing the plate's thickness $t$, and reducing its hole thickness $t_h$, provide generally better net STL gain performance, but at the expense of moving the attenuation band to higher frequency ranges.~It is noticeable that the parameter $a_h$, which controls the radii of the connecting junctions between the holes in the plate portion (for a periodic system), is shown to have a large influence on the coupling-resonance mechanism.~The role of these connecting junctions is analogous to that of the connecting spring between cells in the double-periodicity supercell model in Fig.~\ref{Fig1}d.~As informed by the analytical model (see SM section), reducing this spring's stiffness improves the coupling between the resonance frequencies.~This effect extends to the HP${}^2$ system when increasing the $a_h$ value (for $a_h<16$ mm, the double peaks are lost with just one peak appearing in the attenuation band). \\
\indent~The unique STL double-peak phenomenon may appear also at lower frequency ranges.~This is realized by (1) increasing the resonating mass (e.g., by relaxing the single-material requirement and adding a metal cap on the pillar's tip and adding a second metal disc on the plate's front face) and/or (2) tweaking the geometric parameters (typically by increasing the overall size) of the HP${}^2$ configuration.~The green curve in Fig.~\ref{Fig3} shows an example in which lead discs have been added with a 9~mm increase of the overall thickness, producing enhanced STL response at frequencies around 1~kHz (see the SM section for more details).

\appendix

\setcounter{figure}{0}
\setcounter{equation}{0} 
\setcounter{section}{0}

\renewcommand{\thefigure}{A\arabic{figure}}
\renewcommand{\thetable}{A\arabic{table}}
\renewcommand{\theequation}{A\arabic{equation}}

\vskip 0.15in
\centerline{\textbf{APPENDIX}}
%\textbf{APPENDIX}
%\section{}
\begin{center}
\textbf{Analytical model for the double-periodicity mass-in-mass chain}
\end{center}
%\centerline{\textbf{Analytical model for the double-periodicity mass-in-mass chain}}\\
%\section{\label{sec:AppA} Analytical model for the double-periodicity mass-in-mass chain}
%\noindent \textbf{Analytical model for the double-periodicity mass-in-mass chain}\\
\indent The equations of motion for the outer masses (denoted $M_0$ in Fig.~1d of the main article) of the $k$-th supercell in an infinite chain are given by
\begin{multline}
    M_0\ddot{u}_1^{(k)} + K_0 (2u_1^{(k)} - u_2^{(k)} - u_2^{(k-1)}) + {} \\ {} + k_1 (u_1^{(k)} - u_1^{\prime(k)}) = 0,\label{eq_motion1}
\end{multline}
\begin{multline}
    M_0\ddot{u}_2^{(k)} + K_0 (2u_2^{(k)} - u_1^{(k)} - u_1^{(k+1)}) + {} \\ {} + k_2 (u_2^{(k)} - u_2^{\prime(k)}) = 0,\label{eq_motion2}
\end{multline}
where $K_0$, $k_1$ and $k_2$ are spring parameters defined in Fig.~1d of the main article, $u_i^{(k)}$ is the displacement of each unit $i=\{1,2\}$ in the supercell $k$, and $u_i^{\prime(k)}$ is the displacement of the corresponding unit's internal mass.~The symbol $\ddot{(\bullet)}$ refers to the second time derivative of $(\bullet)$.\\
\indent For the internal masses (denoted $m_1$ and $m_2$ in Fig.~1d of the main article), the equations of motion are
\begin{equation}
    m_i \ddot{u}_i^{\prime(k)} + k_i (u_i^{\prime(k)} - u_i^{(k)}) = 0,\label{eq_internal}
\end{equation}
for $i=\{1,2\}$.\\
\indent According to Floquet's theory, the displacements can be expressed in terms of a frequency $\omega$ and wavenumber $\kappa$ as
\begin{equation}
    u_i^{(k)} = U_i e^{\text{i}(\kappa x_i^{(k)} - \omega t)}, \ \ u_i^{\prime(k)} = U_i^{\prime} e^{\text{i}(\kappa x_i^{(k)} - \omega t)},\label{eq_floquet}
\end{equation}
for $i=\{1,2\}$.~Now, assuming the separation between cells 1 and 2 in each supercell is $a/2$ (with $a$ being the periodicity distance, i.e., the size of the supercell), one can express the position of the second cell, $x_2^{(k)}$, in terms of the first one, $x_1^{(k)}$, as
\begin{equation}
    x_2^{(k)} = x_1^{(k)} + a/2,\label{eq_x2}
\end{equation}
and the positions of the neighbouring cells can be expressed as
\begin{equation}
    x_i^{(k-1)} = x_i^{(k)}-a, \ \ x_i^{(k+1)} = x_i^{(k)}+a\label{eq_xk}
\end{equation}
for $i=\{1,2\}$.\\
\indent Upon introducing Eq.~\eqref{eq_floquet} into Eq.~\eqref{eq_internal}, then
\begin{equation}
    U'_i = \dfrac{k_i}{k_i-\omega^2 m_i} U_i \ \rightarrow \ U'_i = \dfrac{\omega_i^2}{\omega_i^2-\omega^2} U_i,\label{eq_internal2}
\end{equation}
where $\omega_i^2 = k_i/m_i$.~Similarly, considering expressions~\eqref{eq_x2} and \eqref{eq_xk} in Eq.~\eqref{eq_floquet}, and introducing those, along with Eq.~\eqref{eq_internal2}, into Eqs.~\eqref{eq_motion1} and \eqref{eq_motion2}, the resulting system yields
\begin{equation}
    \begin{bmatrix}
    \lambda_1 & \lambda \\
    \lambda & \lambda_2
    \end{bmatrix} 
    \begin{bmatrix}
    U_1 \\
    U_2
    \end{bmatrix} = 
    \begin{bmatrix}
    0 \\
    0
    \end{bmatrix}, \label{eq_sys}
\end{equation}
with
\begin{align}
    &\lambda_i = 2 - \left(\dfrac{\omega}{\Omega_0}\right)^2 \left( 1 + \dfrac{\omega_i^2}{\omega_i^2 - \omega^2}\alpha_i \right),\label{eq_lambda}\\
    &\lambda = e^{\text{i}\kappa a/2} + e^{-\text{i}\kappa a/2},
\end{align}
where $\Omega_0 = K_0/M_0$ and $\alpha_i = m_i/M_0$.\\
\indent For a given frequency, $\lambda_1$ and $\lambda_2$ are determined by Eq.~\eqref{eq_lambda}, and the dispersion relation for system \eqref{eq_sys} is then be obtained by
\begin{equation}
    \lambda_1 \lambda_2 - \lambda^2 = 0,
\end{equation}
or, equivalently,
\begin{equation}
    \lambda_1 \lambda_2 - e^{\text{i}\kappa a} - e^{-\text{i}\kappa a} - 2 = 0.\label{eq_disp}
\end{equation}
From multiplying Eq.~\eqref{eq_disp} by $e^{\text{i}\kappa a}$, and with the change of variable $\eta = e^{\text{i}\kappa a}$, one finds
\begin{equation}
    \eta^2 - (\lambda_1 \lambda_2 - 2)\eta + 1 = 0. \label{eq_quad}
\end{equation}
Once $\eta$ is obtained from Eq.~\eqref{eq_quad}, the real and imaginary parts of the wavenumber, $\kappa_\text{R}$ and $\kappa_\text{I}$ respectively, can be expressed as
\begin{align}
    &\kappa_\text{R}a = \tan^{-1}\left(\dfrac{\eta_\text{I}}{\eta_\text{R}}\right),\\
    &\kappa_\text{I}a = -\ln{\sqrt{\eta_\text{R}^2 + \eta_\text{I}^2}},
\end{align}
where $\eta_\text{R}$ and $\eta_\text{I}$ refer to the real and imaginary parts of $\eta$, respectively.\\
%\begin{equation}
%    4 \cos^2(\kappa a/2) = \lambda_1 \lambda_2 \ \rightarrow \ \kappa a = \cos^{-1} \left(\dfrac{\sqrt{\lambda_1 \lambda_2}}{2}\right).\label{eq_disp}
%\end{equation}
%\indent It should be noticed, from Eq.~\eqref{eq_disp}, that $\kappa$ is real-valued as long as $\lambda_1 \lambda_2 \geq 0$, but it can be complex-valued for $\lambda_1 \lambda_2 < 0$.\\

\begin{figure} [b!]
	\includegraphics{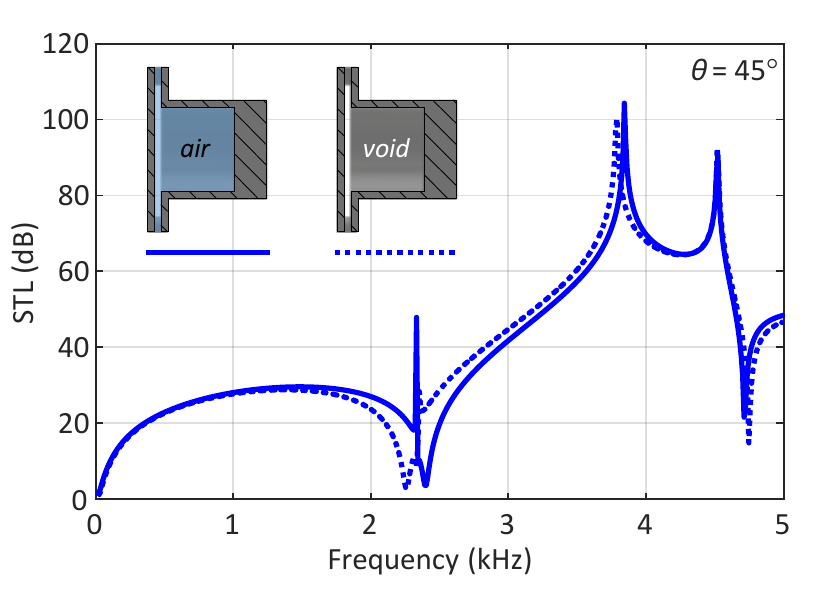}%
	\caption{Comparison of STL curve for the HP${}^2$ configuration and an angle of incidence of 45\textdegree~with the same case but with vacuum instead of air inside the hollow cavities in the plate and  pillar portions.~The difference between both curves is quantified as less than 10\% in STL and is limited to small translation in frequency. \label{FigS1}}
\end{figure}

\begin{center}
\textbf{Effect of air inside the holes}
\end{center}
%\section{\label{sec:AppB} Effect of air inside the holes}
%\noindent \textbf{Effect of air inside the hole}\\
\indent Here we show that the air inside the hollow cavity of the HP${}^2$ panel configuration only has a modest effect on the coupled-resonances mechanism. This is done by comparing results for the nominal  configuration considered in Figs. 3 and 4 in the main article but considering vacuum instead of air in the hollow domain.~Fig.~\ref{FigS1} shows both STL curves exhibiting the same double-peak shape (with just slight differences in the peaks’ locations in the frequency domain), hence verifying that the coupled resonance phenomenon observed, and the overall dynamics of the HP${}^2$ configuration, are not affected by the presence of air inside the holes.\\

\begin{figure}
	\includegraphics{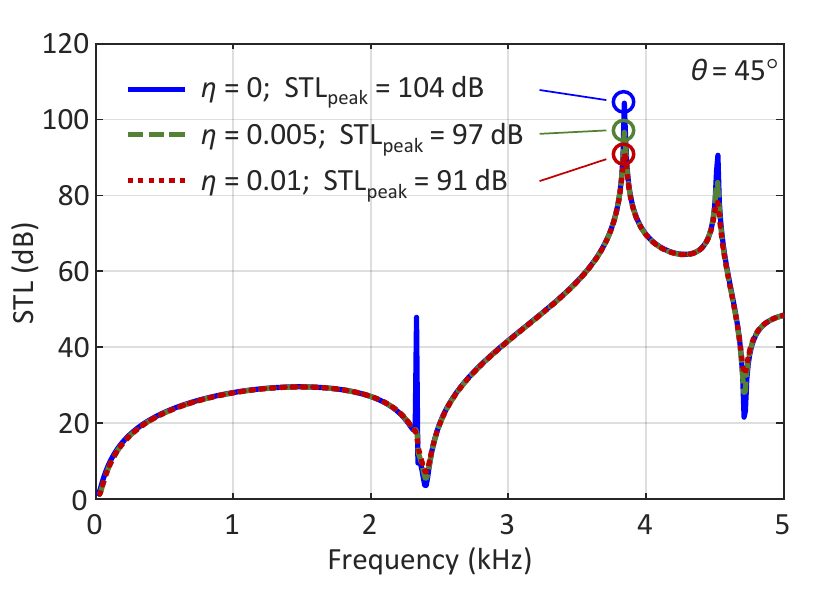}%
	\caption{STL curve for the undamped HP${}^2$ configuration for an angle of incidence of 45\textdegree compared to the same configuration but with isotropic material loss factors of 0.005 and 0.01, respectively, introduced by adding an imaginary part to the Young's modulus of the solid material.\label{FigS2}}
\end{figure}

\begin{center}
\textbf{Effect of loss factor}
\end{center}
%\section{\label{sec:AppC} Effect of loss factor}
%\noindent \textbf{Effect of loss factor}\\
\indent We provide STL curves for the nominal HP${}^2$ configuration but with an isotropic loss factor added to the constitutive material model in order to more accurately represent actual behavior.~The loss factor is introduced through a parameter $\eta$ in the Young's modulus definition $\bar{E}=(1+i\eta)E$, providing it with an imaginary component.~Given the polymeric nature of the material considered (ABS, in this case), a loss factor below 0.01 is considered appropriate \cite{VanBelle2019}.~In Fig.~\ref{FigS2}, it is observed that even with the incorporation of loss for the material considered, the STL peaks still exceed 90~dB.\\

\begin{table*}%[H] add [H] placement to break table across pages
\caption{Geometric parameters for the different HP${}^2$ configurations. Their definitions can be found in Fig.~2 of the main article, except for the metal discs' thickness and diameter, $h_{\text{t}_i}$ and $d_{\text{t}_i}$, which are defined in Fig.~\ref{FigS3}.~All the measures are provided in millimeters (mm).\label{TabS1}}
\begin{ruledtabular}
\begin{tabular}{ccccccccccccc}
Config. & $a$ & $t$ & $h_\text{p}$ & $d_\text{p}$ & $a_\text{h}$ & $t_\text{h}$ & $h_\text{h}$ & $d_\text{h}$ & $h_\text{t1}$ & $d_\text{t1}$ & $h_\text{t2}$ & $d_\text{t2}$ \\ \hline
D     & 25 & 4 & 14 & 15 & 24 & 0.5 & 12 & 11.5 & -   & -  & -   & -  \\
D$^\prime$  & 50 & 8 & 14 & 30 & 49 & 4.5 & 12 & 26.5 & 3.5 & 30 & 1.5 & 10 \\
D$^{\prime\prime}$ & 50 & 8 & 14 & 30 & 49 & 4.5 & 12 & 26.5 & 3.5 & 30 & -   & -  \\
\end{tabular}
\end{ruledtabular}
\end{table*}

\begin{figure}
	\includegraphics{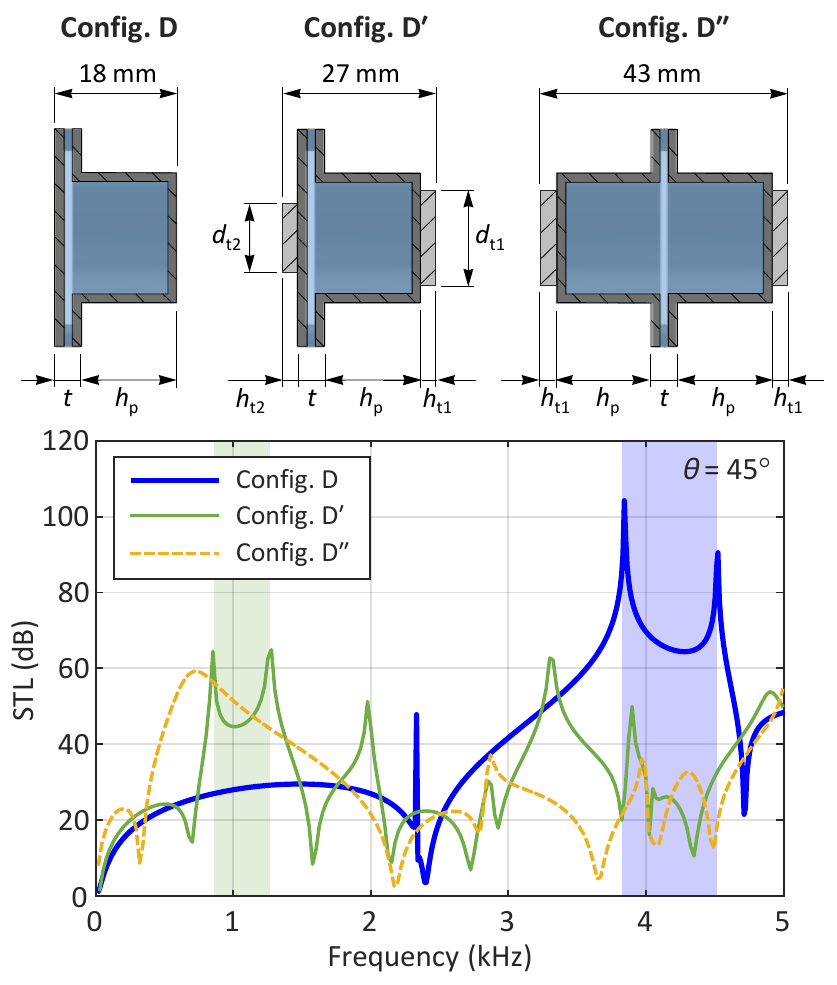}%
	\caption{~STL responses of different variations of the HP${}^2$ design (Configs.~D$^\prime$ and D$^{\prime\prime}$) compared to the original design presented in the main article (Config.~D). For Config.~D$^\prime$, thin lead discs of diameters $d_{\text{t1}}$ and $d_{\text{t2}}$ and thicknesses $h_{\text{t1}}$ and $h_{\text{t2}}$ are added to the pillar's tip and plate's front face, respectively.~For Config.~D$^{\prime\prime}$, the same hollow pillar with the lead disc attached is symmetrically replicated on the plate's front face.~The shaded areas correspond to frequency ranges between double peaks caused by the coupled resonances effect.\label{FigS3}}
\end{figure}

\begin{center}
\textbf{Target frequency range}
\end{center}
%\section{\label{sec:AppC} Target frequency range}
%\noindent \textbf{Target frequency range}\\
\indent In the main article, most results shown are for a specific configuration exhibiting enhanced STL performance in the mid-frequency range (i.e., between 3 and 5~kHz).~This is primarily motivated by our interest in focusing on the modal dynamics of the HP${}^2$ concept and the emerging coupled resonance effect.~However, it is desirable in practice to target the low-frequency regime (i.e., around 1~kHz or lower) given the characteristics of the human ear's auditory perception \cite{Rossing2007}.~In Fig.~\ref{FigS3}, we show that the same coupled resonance effects are also observed in such frequency ranges by appropriately adapting the HP${}^2$ dimensions, geometric parameters, and material properties.~In order to reach the sub-1 KHz regime, either more compliant rubber-like materials may be used (at the expense of losing some load-bearing capacity), or masses may be added on the resonating elements of a given design (in this case probably sacrificing the overall thickness).\\
\indent For the example shown in Fig.~\ref{FigS3}, thin discs made of lead ($E=14$~GPa, $\rho=11600$~kg/m${}^3$, and $\nu=0.42$ and) have been added to the pillar's tip and on the plate's front face to obtain an enhanced STL response via the coupled resonance effect in a frequency range around 1~kHz.~The geometric parameters for this enhanced HP${}^2$ configuration are given in Table~\ref{TabS1}.~While the unit cell's size $a$ and pillar's diameter $d_p$ are each doubled, in the thickness direction only 4~mm in $t_\text{h}$ and the two lead discs' combined thickness of $h_{\text{t1}}+h{_\text{t2}}=5$~mm have been added; this amounts to a 50~\% increase with respect to the overall thickness of the nominal configuration.\\
\indent In Fig.~\ref{FigS3}, we show also the STL response for a double-pillar configuration in which the same holed pillar is replicated on the plate's front face.~Interestingly, even though the coupled resonance effect is lost and the overall thickness is largely increased (in this case, an additional 25~mm), the panel exhibits a decent boost in the STL response at even lower frequencies.

\vskip 0.15in
\centerline{\textbf{ACKNOWLEDGEMENT}}

David Roca gratefully acknowledges the support received by the Spanish Ministry of Education through the FPU program for PhD grants (FPU16/06113).

\bibliography{bibfile}

\end{document}